# Role of critical spin fluctuations in ultrafast demagnetization of transition-metal rare-earth alloys.

V. López-Flores[1], N. Bergeard[1], V. Halté[1], C Stamm[2], N. Pontius[2], M. Hehn[3], E. Otero[4], E. Beaurepaire[1], C. Boeglin[1]

[1]*Institut de Physique et de Chimie des Matériaux de Strasbourg and NIE, UMR7504, CNRS et Université de Strasbourg, 67034 Strasbourg, France*

[2]*Institut für Methoden und Instrumentierung der Forschung mit Synchrotronstrahlung Helmholtz-Zentrum Berlin für Materialien und Energie GmbH, Albert-Einstein-Str. 15, 12489 Berlin, Germany*

[3]*Institut Jean Lamour, Université Henri Poincaré, Nancy, France*

[4]*Synchrotron SOLEIL, L'Orme des Merisiers, Saint-Aubin, 91192 Gif-sur-Yvette, France.*

Ultrafast magnetization dynamics induced by femtosecond laser pulses have been measured in ferrimagnetic $Co_{0.8}Gd_{0.2}$, $Co_{.74}Tb_{.26}$ and $Co_{.86}Tb_{.14}$ alloys. Using element sensitivity of X-ray magnetic circular dichroism at the Co $L_3$, Tb $M_5$ and Gd $M_5$ edges we evidence that the demagnetization dynamics is element dependent. We show that a thermalization time as fast as 280 ±30 fs is observed for the rare-earth in the alloy, when the laser excited state temperature is below the compensation temperature. It is limited to 500 ±100 fs when the laser excited state temperature is below the Curie temperature (Tc). We propose critical spin fluctuations in the vicinity of $T_C$ as the mechanism which reduces the demagnetization rates of the 4f electrons in transition-metal rare-earth alloys whereas at any different temperature the limited demagnetization rates could be avoided.

# I. INTRODUCTION

The application of ultra-short laser pulses allows the manipulation of the magnetization of films down to the femtosecond time scale. This is relevant for technological applications such as magnetic recording and data storage, but also addresses interesting fundamental problems. Therefore it is essential to explore the different fundamental processes taking place during the first hundred femtoseconds and related to the interaction between laser light and matter. The first observation of ultrafast demagnetization in Ni films, identifying electron, spin, and lattice thermalization phenomena, has led to numerous attempts to explore the ultrafast dynamics in various transition-metals (TM) [1-8]. The standard technique used in these studies is time-resolved magneto-optic Kerr effect (TR-MOKE) but in the last few years time-resolved X-ray magnetic circular dichroism (TR-XMCD) with femtosecond time resolution has been competing with the all optical techniques [9-12]. Despite these efforts, fundamental questions related to the conservation of the angular momentum are still open.

In the case of Rare Earths (RE) metals, demagnetization mechanisms involving spin-flip processes were proposed to explain the ultrafast demagnetization [12]. Thin Gd and Tb films were studied by Wietstruk et al. [12] using TR-XMCD and the results suggest that demagnetization evolves along two different regimes, at picosecond and femtosecond time scales. The existence of a large orbital moment in Tb (L=3) compared to L=0 in Gd has been invoked to explain the stronger spin-lattice interaction in Tb which could in turn explain a faster spin-lattice relaxation in the ps time scale (~8ps for Tb and ~40ps for Gd). Moreover, in the femtosecond time range, the ultrafast demagnetization process shows a relatively long thermalization time of $\tau_{ther} = 750$ fs, much larger than in most of the transition metals where $\tau_{ther} \sim 100 - 300$ fs [5, 6, 8, 9,10]. Surprisingly, both bulk RE elements show similar sub-picosecond characteristic times irrespective of their orbital moments. Recently, in ferrimagnetic $Fe_{.75}Co_{.34}Gd_{.22}$ it was shown that Gd shows a demagnetization time of 430 fs whereas Fe is much faster [11]. The numerical values of the thermalization times have not been explicitly addressed. Therefore one could have been speculating that it is the 3d-5d hybridization in TM-RE alloys which speeds up the 4f ultrafast dynamics with respect to bulk RE systems. In order to resolve such questions, more studies on specific ferrimagnetic materials, consisting of two antiparallel-coupled sublattices, are requested. Element-specific TR-XMCD allows for an independent characterization of both magnetization dynamics. Such ferrimagnetic systems may also possess a magnetic compensation temperature ($T_{comp}$) where the magnetic moments in both sublattices are equal in amplitude and cancel each other [13]. Therefore we have decided to perform XMCD measurements for Co-RE alloys (RE=Gd,Tb).

We show that the long thermalization times previously observed for RE elements [11, 12] can be accelerated by a factor 2 when the laser excited state temperature is below $T_{comp}$ instead of $T_C$. To this end, we designed specific $Co_{1-x}Gd_x$ and $Co_{1-x}Tb_x$ alloys where the compensation temperature is either below or far above the working temperature.

## II. EXPERIMENT

The time resolved XMCD experiments were performed at the femtoslicing beam line of the BESSY II synchrotron radiation source of the Helmholtz-Zentrum Berlin [9, 10]. The experiment requires a pump and probe setup where the short X-ray pulses are synchronized with a femtosecond pump laser (near IR wavelength - 790nm, 3 kHz pump pulse repetition rate). The X-ray pulse duration of about 100 fs, in the femtoslicing operation mode, ensures a global time resolution of ~130 fs (see refs. [9, 10] for details). The magnetization dynamics have been measured by monitoring the transmission of circularly polarized X rays pulses tuned to specific core level absorption edges as a function of a pump-probe delay in the range 0-4 ps. The photon energy was set to the $CoL_3$ and the $TbM_5$ or $GdM_5$ edges using the Bragg Fresnel reflection zone plate monochromator UE56/1-ZPM. The pump fluences used during our experiment were adjusted to 8 mJ/cm$^2$ for $Co_{0.8}Gd_{0.2}$, 12 mJ/cm$^2$ for $Co_{0.74}Tb_{0.26}$ and 21 mJ/cm$^2$ for $Co_{0.86}Tb_{0.14}$ in order to reach close amplitudes of demagnetization at the $CoL_3$ edge (~ 60%). To ensure magnetic saturation in the films a magnetic field of ± 5 kOe is applied during the pump-probe and during the static XMCD measurements. Static XMCD measurements were recorded at the beam line DEIMOS of the French synchrotron facility SOLEIL and performed in transmission, using the same geometry as the time-resolved experiments. The static XMCD measurements shown in Fig. 1 were performed for temperatures corresponding to the working conditions during the time resolved XMCD (290K for $Co_{0.8}Gd_{0.2}$, and 400K for $Co_{0.74}Tb_{0.26}$ and $Co_{0.86}Tb_{0.14}$). The $Co_{0.8}Gd_{0.2}$ alloy shows in-plane anisotropy whereas both $Co_{1-x}Tb_x$ alloys show out-of-plane anisotropy as evidenced by XMCD magnetometry (see inset in Fig. 1). Therefore, both CoTb films were measured along the surface normal to the films whereas $Co_{0.8}Gd_{0.2}$ was measured at 35 degrees from the normal to the surface plane.

The 15 nm Co-RE alloys have been grown on $Si_3N_4$ membranes. The films are characterized by a compensation temperature $T_{comp}$, where the magnetic moments of the Co and the rare-earth sublattices compensate and by a specific Curie temperature $T_C$ where the magnetic order is lost. The amorphous $Co_{0.8}Gd_{0.2}$, $Co_{.74}Tb_{.26}$ and $Co_{.86}Tb_{.14}$ alloy films were optimized in order to obtain moderate saturation fields of 5 kOe or less, compatible with the magnetic field available in the experiment. Furthermore, the films were selected for their variation in compensation temperatures ($T_{comp}$ = 150K, 550K and < 40 K) and Curie temperatures (resp. $T_C$ = 450K, 650K and 750K). In Fig. 2 we plot the qualitative temperature dependent XMCD (T) amplitudes at the $L_3$ edge of Co and $M_5$ edge of RE as a function of the temperature of the samples and we illustrate the relative positions of $T_{comp}$ and $T_C$. This plot is extracted from static XMCD measurements in the temperatures range 4K - 400K and from the literature for higher temperatures [13]. The pump-probe experiment was performed at film temperatures close to room temperature or slightly above (large filled circles in Fig. 2) so that the laser excited state temperature is in the vicinity of $T_C$ or $T_{comp}$ (small filled circles). The large blue (red) circles in Fig. 2 show the XMCD (T) for the RE (resp. TM) elements taking into account the permanent heating induced by the 3 kHz laser. The small filled circles show the laser excited state temperature after the thermalization of the electrons when hot electrons are present. This defines the thermalization of spins. The parameter $\tau_{ther}$ can then be related to the excited state temperature. It illustrates the different temperature ranges in which we performed the pump probe experiments for all our compounds. From a more general stand point, one has to be cautious when drawing parallels between ultrafast dynamics and the temperature dependent magnetization curve.

### III. RESULTS

In Fig. 1 we show the results of static XMCD measurements for our alloys. The ferrimagnetic coupling aligns the Co magnetic moments opposite to the RE moments, and for each element the absolute orientation in $Co_{0.8}Gd_{0.2}$ and $Co_{0.86}Tb_{0.14}$ is opposite to the one of $Co_{0.74}Tb_{0.26}$. This is because the compensation temperatures are either below or above the sample temperature during the measurements ($T_{comp}$ ($Co_{0.8}Gd_{0.2}$) =150 K, $T_{comp}$ ($Co_{0.86}Tb_{0.14}$) < 40 K and $T_{comp}$ ($Co_{0.74}Tb_{0.26}$) = 550 K). Measuring the XMCD signal (insets) at the $CoL_3$ and Gd (or Tb) $M_5$ edges as a function of the magnetic field confirms that magnetic saturation is reached before 5 kOe around room temperature. We apply the XMCD sum rules [14, 15] on both $CoL_{2,3}$ and Gd (resp. Tb) $M_{4,5}$ edges in order to define the individual quantitative effective magnetic moments $S^{eff}$ (the effective spin moments includes the contribution from the magnetic dipole operator Tz) and L in absolute units (ℏ/at). The different magnetic moments measured at room temperature and evaluated as explained below are listed on Fig. 1. In $Co_{0.8}Gd_{0.2}$ the effective spin moment of Co is $S^{eff}$(Co) = 0.46 ± 0.1 ℏ/at whereas the orbital moments is L(Co) = 0.12 ± 0.03 ℏ/at. In $Co_{0.74}Tb_{0.26}$ $S^{eff}$(Co) = -0.45 ± 0.1 ℏ/at whereas the orbital moments is L(Co) = -0.11± 0.03 ℏ/at. In comparison larger magnetic moments (Seff(Co)=0.72 ℏ/at and L(Co)=0.15 ℏ/at) are measured for $Co_{0.86}Tb_{0.14}$ related to the higher Co concentration. Quantitatively one notices that large $L/S^{eff}$ ratios of 0.21- 0.29 are obtained for the Co atoms in all alloys whereas bulk Co shows a ratio of 0.13 [15] suggesting that large magnetocrystalline anisotropy energies (MAE) are present in all Co-RE compounds through a significant contribution of the 3d-5d hybridization [16, 17].

For Co where L is small and for Gd where L is zero we can neglect the magnetic dipole term Tz, so that for those elements $S^{eff}$ = S. In the case of Tb the contribution of Tz cannot be neglected so that the spin moment S can only be derived using a correction factor of $(1+3Tz/S)^{-1}$ =3/2 [15, 18]. Finally, for $Co_{0.74}Tb_{0.26}$ one obtains S(Tb) = 1.35 ± 0.2 ℏ/at confirming a ratio L/S ~1 for Tb similar to previous results [18]. For $Co_{0.86}Tb_{0.14}$ we have S(Tb) = 2.01 ± 0.2 ℏ/at and L(Tb) = 2.10 ± 0.2 ℏ/at, leading to the same result (L/S ~1). The values measured for the RE elements are significantly lower than what is expected from Hund's rules (S(Gd) = 7/2, L(Gd) = 0, S(Tb)= 3, L(Tb) = 3). This is due to the structural disorder and finite temperature effects.

These data also define the ratio between the Curie temperature $T_C$ and the total magnetic moment M ($T_C$ /M) which in the model proposed by Koopmans et al. [19] is predicted to be the scaling factor of the demagnetization rate. Extending this model to ferrimagnets [20], the ratios obtained for $Gd_{0.2}$ ~80 K/$\mu_B$ and $Tb_{0.26}$ ~160 K/$\mu_B$ would predict that $Co_{0.74}Tb_{0.26}$ exhibits a slower 4f demagnetization (larger $\tau_{ther}$) than $Co_{0.8}Gd_{0.2}$. This is however not what is observed in our experiments. The main difference, besides the change of the material, is the fact that we do not reach the same laser excited state temperature for both systems.

Figure 3a shows the ultrafast dynamics obtained during the pump-probe experiment at the $CoL_3$ and $TbM_5$ edges for the $Co_{0.74}Tb_{0.26}$ alloy. The low XMCD amplitudes measured during the slicing experiment are re-normalized to the quantitative static XMCD measurements (Fig. 1). For instance in figure 3a for $TbM_5$, the y-axis corresponds at t < t0 to the static normalized XMCD amplitude measured at the $TbM_5$ edge at DEIMOS beam line. This quantitative values are used as the initial values in figure 3 and 4 (delays t < t0) and lead to the correct relative XMCD between Co $L_3$ , Tb $M_5$ and Gd $M_5$.

As a means for comparison in Figure 3b we show the ultrafast dynamics obtained at the $CoL_3$ and $GdM_5$ edges for the $Co_{0.8}Gd_{0.2}$ alloy. In order to verify that no inversion of the XMCD sign occurs during the demagnetization of the films, we have performed longer delay scans (not shown) at the Tb and Gd $M_5$ edges which have confirmed that the temperature did not reach $T_{comp}$ in $Co_{0.74}Tb_{0.26}$. This is an important difference with the experiments in [11] where the spin switching mechanism was addressed. The analysis of demagnetization dynamics is carried out using the same formalism as in Boeglin et al. [10] where a double exponential function has been used to fit the XMCD pump-probe data. The simulations are obtained by approximating the rate equations of the two temperature model, using $\tau_{ther}$ and $\tau_{s-ph}$ as fit parameters defined respectively as the thermalization time and the relaxation time for the spins to other degrees of freedom (lattice, external bath). Surprisingly lower thermalization times are evidenced in the $Co_{0.74}Tb_{0.26}$ compound, namely $\tau_{ther} = 180 \pm 40$ fs for Co and $\tau_{ther} = 280 \pm 30$ fs for Tb. The thermalization times in the $Co_{0.8}Gd_{0.2}$ film are $\tau_{ther} = 200 \pm 30$ fs for Co and $\tau_{ther} = 480 \pm 40$ fs for Gd. We notice that the demagnetization time obtained for Gd $M_5$ in $Co_{0.8}Gd_{0.2}$ is very close to published values obtained for Gd in $Fe_{.75}Co_{3.4}Gd_{.22}$ [11].

Interestingly, in Fig. 3 we find the ultrafast demagnetization dynamics to be significantly faster in Tb ($\tau_{ther} = 280 \pm 30$ fs) than in Gd ($480 \pm 40$ fs). This is unexpected since Wietstruk et al. reported a two step demagnetisation for pure Gd and Tb [12] with similar "fast" processes in the femtosecond range ($\tau_{ther}$ ~750 fs) involving hot electrons. We show here that alloying RE with Co leads to different ultrafast demagnetization dynamics in the "fast" process, where an acceleration of the demagnetization rate by nearly a factor of 2 is observed for Tb compared to Gd. Moreover, using the chemical selectivity of XMCD, we verify that the thermalization time of Co is very close in both alloys, (~$\tau_{ther} = 200 \pm 40$ fs for Co). The relative acceleration of the thermalization for RE compared to the Co 3d element raises intriguing questions here.

In the case of transition metals, it is known that the demagnetization processes occur during the thermalization of the excited hot electrons [4] and that the corresponding rate of demagnetization depends on the absorbed laser energy and on the detailed dynamics of the spin-resolved band structure [6, 19, 21]. The fact that in both $Co_{.76}Tb_{.26}$ and $Co_{0.8}Gd_{0.2}$ alloys we found very similar thermalization times for Co and also the same demagnetization amplitude is related to the similar experimental pumping conditions. What is new and striking is that different thermalization times are evidenced for the Tb and Gd constituents, showing no simple correlation between the thermalization times in RE and the absorbed laser energies. We propose that the observed acceleration of the thermalization in Tb is related to the fact that the laser excited state temperature is in the vicinity of $T_C$ in the case of $Co_{0.8}Gd_{0.2}$ while it is in the vicinity and below $T_{comp}$ in the case of $Co_{0.74}Tb_{0.26}$.

In order to confirm the correlation between the proximity to $T_C$ and a slowing down of the ultrafast dynamics of TM-RE compounds, we probed a different $Co_xTb_{1-x}$ compound for which $T_C$ can be reached by laser heating while keeping the fluence bellow the threshold damage. This also serves to exclude the nature and the exchange coupling of the rare-earth constituent as cause for the different demagnetization times. In Fig. 4 we show the ultrafast dynamics at Tb $M_5$ edges in $Co_{0.86}Tb_{0.14}$ and in $Co_{0.74}Tb_{0.26}$. The analysis in both alloys determines that $\tau_{ther}$ = 500 ± 100 fs in $Co_{0.86}Tb_{0.14}$ and $\tau_{ther}$ = 280 ± 30 fs in $Co_{0.74}Tb_{0.26}$. Although the statistics of the data are not of equal quality, they allow us to confirm the relevance of the nature of the transition toward which the system is pumped. Moreover, these results exclude any proportionality between the demagnetization time and the Co concentration in CoTb alloys as suggested by simple 3d-5d hybridization arguments (bulk Tb, $Co_{.76}Tb_{.26}$ and $Co_{.86}Tb_{.14}$ are defined by resp. $\tau_{ther}$ = 750 fs, 280 fs and 500 fs).

Finally, we will develop a model in which a slower thermalization is obtained when the temperature is increased in the vicinity of Curie temperature $T_C$, provided that one involves localized 4f electrons as the ones being probed. Since ultrafast dynamics are considered, one cannot assume a quasi static behavior in our system in the femtosecond regime. However, by analogy with phase transition models, we assume that the system of localized 4f electrons approaches equilibrium with characteristic times $\tau_{relax}$ related to the magnetization dynamic M(t) by :

$$M(t) = M(0) \exp[- t/ \tau_{relax}], \qquad (1)$$

where

$$\tau_{relax} \sim | (T-T_c)|^{-z\nu} \qquad . \qquad (2)$$

The critical exponent $z\nu$ defines the divergence of the spin relaxation and is model dependent so that in the vicinity of the critical point $T_C$, the spin relaxation time increases.

For the present qualitative discussion we expect the critical regime to happen in the temperature range $0.7 \leq T/T_C \leq 1$ in order to cover our experimental situations. In the case of the 3D Ising model, $z\nu \sim 1.3$ which is in reasonable agreement with ultrafast spin dynamics measured on single crystal $Sr_2FeMoO_6$.[22]. In a recent work by Kantner et al. [23], longer demagnetization times have been evidenced in SrRuO3 films when using laser conditions to reach excited state temperature in the vicinity of $T_C$, confirming that these times are inversely proportional to T-Tc. Similarly, in a model including interactions between thermodynamic baths [1], the relaxation time of each bath (in the weak perturbation regime [24]) is proportional to the specific heat, leading to a strong increase as the Curie temperature is approached (since the spin specific heat diverges at $T_C$). Similar calculations, from Manchon et al. [25] discuss the effect of temperature in the ultrafast magnetization dynamics in metals. Therefore we expect an increase of the spin relaxation time $\tau_{relax}$ as $T_C$ is approached. Thus, we propose that the different ultrafast dynamics in RE are related to the different temperatures reached after the thermalization of the electrons. In $Co_{0.74}Tb_{0.26}$ the reached temperature is in the vicinity of $T_{comp}$ = 550K whereas in $Co_{0.8}Gd_{0.2}$ the reached temperature is in the vicinity of $T_C$, ie approaching the critical fluctuations [13]. This model is confirmed in Figure 4 by comparing the ultrafast dynamics in $Co_{0.74}Tb_{0.26}$ and $Co_{0.86}Tb_{0.14}$. Previous work performed using TR-MOKE at a wave length of 400 nm show that in $Fe_{.65}Tb_{.35}$ a "slow" dynamics is present [26] when the laser excited state temperature is in the vicinity of $T_C$ = 380K, confirming our model. The latter experiment is also highly sensitive to the RE magnetization dynamics confirmed by recent results showing that element sensitivity can be achieved using TR-MOKE [27]. Finally, we wish to point out that although the 4f elements studied so fare are demagnetization rate limited close to $T_C$ by the divergence of 4f spin fluctuations, this can be overcome using $T_{comp}$ in the ferrimagnetic alloys.

## IV. DISCUSSION

Let us now discuss the underlying physical mechanism explaining the accelerated 4f electron demagnetization rates in $Co_{.74}Tb_{.26}$. In the theoretical framework of the Elliot-Yafet model [3] the demagnetization rate is interpreted using spin-phonon scattering. The spin-phonon interaction strength is considered to be proportional to the magnetocrystalline anisotropy energy which is large in our Co-RE compounds. In our case, the large spin-orbit coupling in Tb should favour fast thermalization processes via enhanced spin-flip scattering [12]. This is in contradiction with the results from the $Co_{0.86}Tb_{0.14}$ sample where the thermalization is much slower than for $Co_{.74}Tb_{.26}$ even if the spin-orbit coupling is equivalent. Time-resolved XMCD measurements suggested that the angular momentum can be transferred via spin-flip scattering from the spin system to the external bath [9]. It has been suggested that this bath can be identified with the phonons [19]. However, quantitative *Ab Initio* investigations show that the contribution of electron-phonon spin-flip is too small to describe the total femtosecond demagnetization [28]. So we can conclude that differences in the electron-phonon contributions to the demagnetization rates cannot explain the strong acceleration in $Co_{.74}Tb_{.26}$.

Recent magneto-optic measurements performed on FeCoGd alloys by Medapalli et al. [29] suggest that at T < $T_{comp}$ the angular momentum is efficiently transferred from Fe toward Gd during the first few hundred femtoseconds, leading to a faster demagnetization. This is in line with recent theoretical calculations obtained for ferrimagnetic FeGd [30]. Considering the antiparallel alignment between the 3d and 5d atomic moments they conclude that the transfer of angular momentum should imply a faster demagnetization in both elements, RE and TM. Unfortunately, Medapalli et al. [29] could not evidence the effect, related to the lack of chemical sensitivity of the magneto-optic experiment. However, our time-resolved XMCD results, show that the fastest demagnetization ever measured for rare earth elements is obtained at $T < T_{comp}$, which is definitively consistent with the theoretical model [30]. We suggest that fast demagnetization of 4f electrons should be possible at any temperature and that in the vicinity of Tc the demagnetization times are increasing due to critical spin fluctuations. In order to trigger a sizable change in the magnetization we have chosen to use a particular temperature $T_{comp}$ which is an essential ingredient of the TM-RE ferrimagnets. In opposite to Medapalli [29] we do not introduce a particular threshold temperature $T_{comp}$ in order to explain changes in the demagnetization times. Our argumentation is based on the fact that the different thermalization times are related to the critical 4f spin fluctuations near $T_C$, reducing the demagnetization rates, whereas below $T_{comp}$ (i.e. far from Tc) this limitation is absent. Finally, ultrafast demagnetization times close to those observed for 3d elements can be achieved for RE elements in ferrimagnetic alloys [5, 6, 8, 9,10,19].

## V. CONCLUSION

We have studied ultrafast demagnetization dynamics using element-specific XMCD in a time resolved way for different Co-RE alloys where the compensation temperature can be tuned. We found faster demagnetization times for RE when the system is excited toward the compensation temperature rather than the Curie temperature. This is explained by the emergence of critical spin fluctuations near the Curie temperature. In $Co_{0.74}Tb_{0.26}$, the Tb moment exhibits a thermalization time as fast as 280 fs when pumped toward $T_{comp}$. This is the shortest time that has been observed so far in rare-earth elements. This work opens new prospects for the ultrafast magnetization manipulation in suitably tailored TM-RE compounds [31].

Acknowledgments:

We are indebted to A. Eschenlohr, K. Holldack, R. Mitzner and T. Kachel for the help and support during the femtoslicing experiments, P. Ohresser for help and to support during the XMCD measurements at DEIMOS and A. Boeglin for discussions and correction of the manuscript.

This work was supported by the CNRS – PICS, by Université de Strasbourg and the E.U. Contract Integrated Infrastructure Initiative I3 in FP6-Project No. R II 3 CT-2004-5060008, BESSY IA-SFS Access Program, by the "Agence Nationale de la Recherhce" in France via the project EQUIPEX UNION: # ANR-10-EQPX-52" and by the German Ministry of Education and Research BMBF Grant 05K10PG2 (FEMTOSPEX). V.L.-F. acknowledges Ministry of Education of Spain (Programa Nacional de Movilidad de Recursos Humanos del Plan Nacional de I-D+i 2008-2011) for financial support.

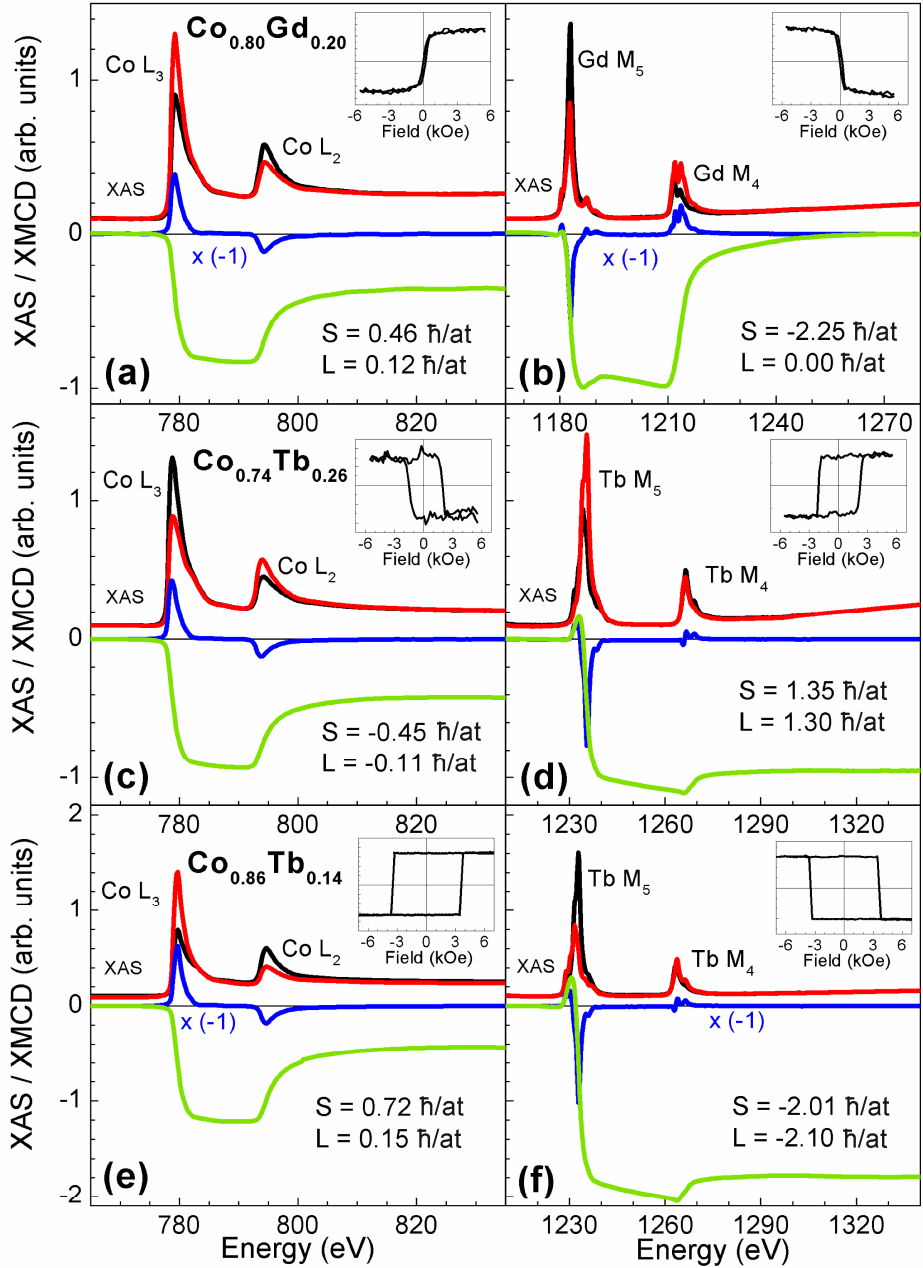

Figure 1: (Color online) Static XMCD obtained at Co $L_{2,3}$ Gd $M_{4,5}$ and Tb $M_{4,5}$ edges using a 0.5 Tesla magnetic field. The red and black lines are the X-ray absorption spectra recorded for parallel and antiparallel orientation of the magnetic field and incoming light helicity. The blue lines are the differences defining the XMCD spectra whereas the green lines are the energy integrated function of the XMCD lines. For clarity the XMCD and the integral functions are multiplied by -1 in fig a), b), e), f). In the inset we show the hysteresis obtained at the different edges (a) Co $L_3$, (b) Gd $M_5$, (c) Co $L_3$ and (d) Tb $M_5$ by cycling a 6 kOe field. The plotted graphs correspond to measurements performed at T= 290 K (a,b) and T= 400 K (c,d,e,f). The error bars obtained by repetitive XMCD measurements are the following : for RE, S and L ± 0.2 ℏ/at and for Co, S ± 0.1 ℏ/at and L ± 0.03 ℏ/at corresponding to a percentage of the numerical values of S and L.

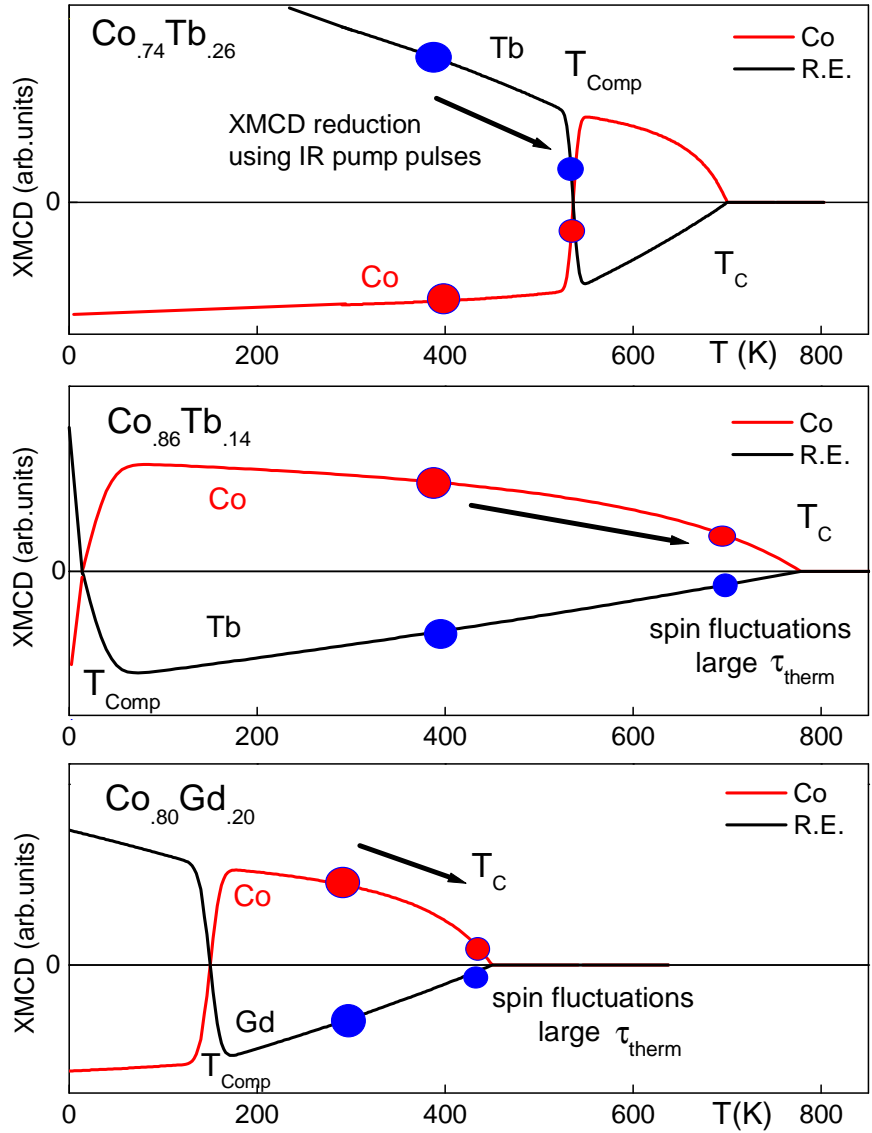

Fig 2. Lopez-Flores et al.

Figure 2: (Color online) Sketch of the magnetization (proportional to the XMCD) versus temperature for $Co_{.74}Tb_{.26}$, $Co_{.86}Tb_{.14}$ and $Co_{0.8}Gd_{0.2}$ with respect of their compensation and Curie temperatures ($T_{comp}$ and $T_C$) [13]. The compensation temperature is either above ($T_{comp}$ = 550K for $Co_{.74}Tb_{.26}$) or below ($T_{comp}$ = 10K - 150K for $Co_{.86}Tb_{.14}$ - $Co_{0.8}Gd_{0.2}$) the sample temperature during the pump-probe experiment (the position is marked by the large filled circles). The large blue and red circles show the XMCD amplitudes obtained for the RE and resp. TM elements in the compounds at the working temperatures, taking into account the permanent heating by the IR laser. The small filled circles show the excited state temperature after the thermalization of the electrons. This position is characterized by the minimum in the demagnetization curves observed for the TR-XMCD signal. The different pump fluences used during our experiment were adjusted to 12 mJ/cm$^2$ for $Co_{0.74}Tb_{0.26}$, 21 mJ/cm$^2$ for $Co_{0.86}Tb_{0.14}$ and 8 mJ/cm$^2$ for $Co_{0.8}Gd_{0.2}$, and is proportional to the increase in temperature (after the thermalization of the electrons) illustrated by the black arrows.

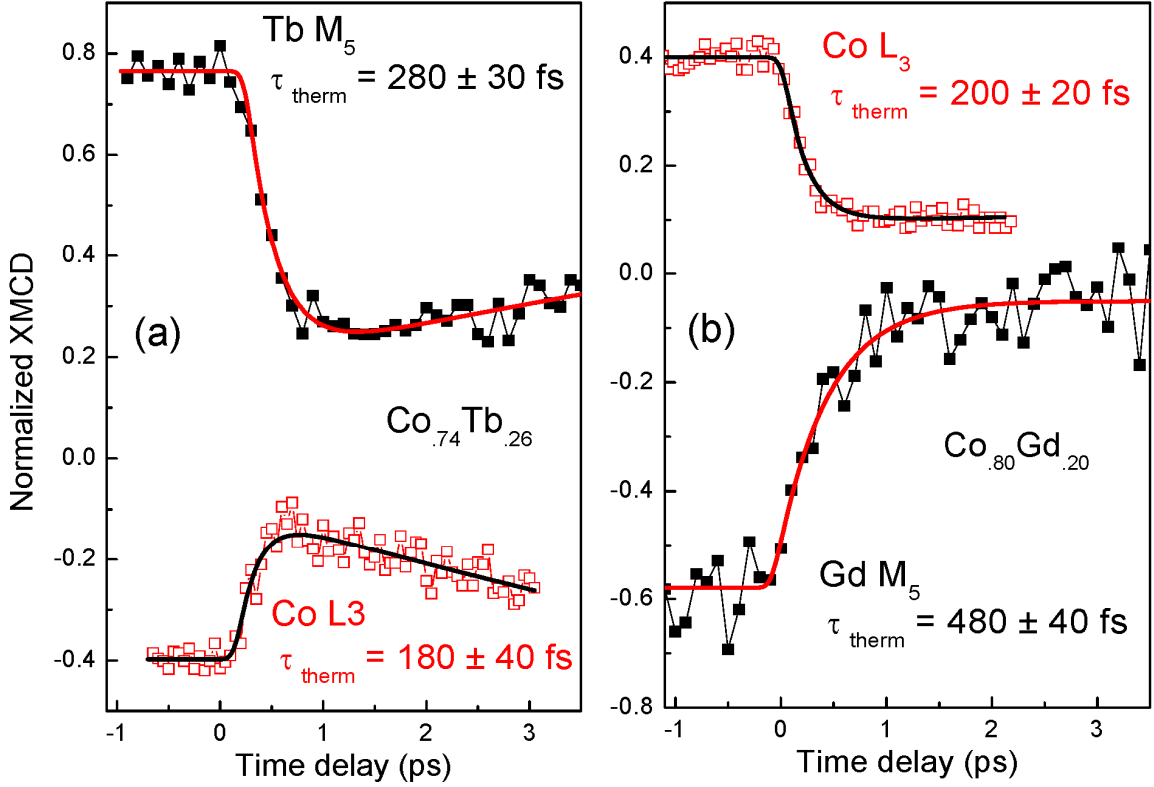

Figure 3: (Color online) Ultrafast dynamics of (a) Co $L_3$, Tb $M_5$ for the $Co_{.74}Tb_{.26}$ alloy and (b) Co $L_3$, Gd $M_5$ edges for the $Co_{.80}Gd_{.20}$ alloy. The respective simulations obtained by approximating the rate equations of the two temperature model are superposed as smooth lines. For the sample $Co_{.74}Tb_{.26}$ the best fits are obtained for $\tau_{therm}$ = 280 ± 30 fs at the Tb $M_5$ edge and $\tau_{therm}$ = 180 ± 40 fs at the Co$L_3$ edge. For the sample $Co_{.80}Gd_{.20}$ the best fits are obtained for $\tau_{therm}$ = 480 ± 40 fs at the Gd$M_5$ edge and $\tau_{therm}$ = 200 ± 30 fs at the Co$L_3$ edge. The XMCD values at the working temperature (t < $t_0$) are provided by using the XMCD data at the Co $L_3$, Tb$M_5$ and Gd$M_5$ edges from the static measurements as given in Fig.1. The working temperature of the sample $Co_{.74}Tb_{.26}$ is T= 400K whereas $T_{comp}$ = 550K. The working temperature of $Co_{.80}Gd_{.20}$ is T= 290K whereas $T_C$ = 450K. During the pump-probe experiments close values of laser fluences were used (8-12 mJ/cm$^2$), corresponding to close temperature intervals between working temperatures and the laser excited state temperatures where ultrafast demagnetization is observed.

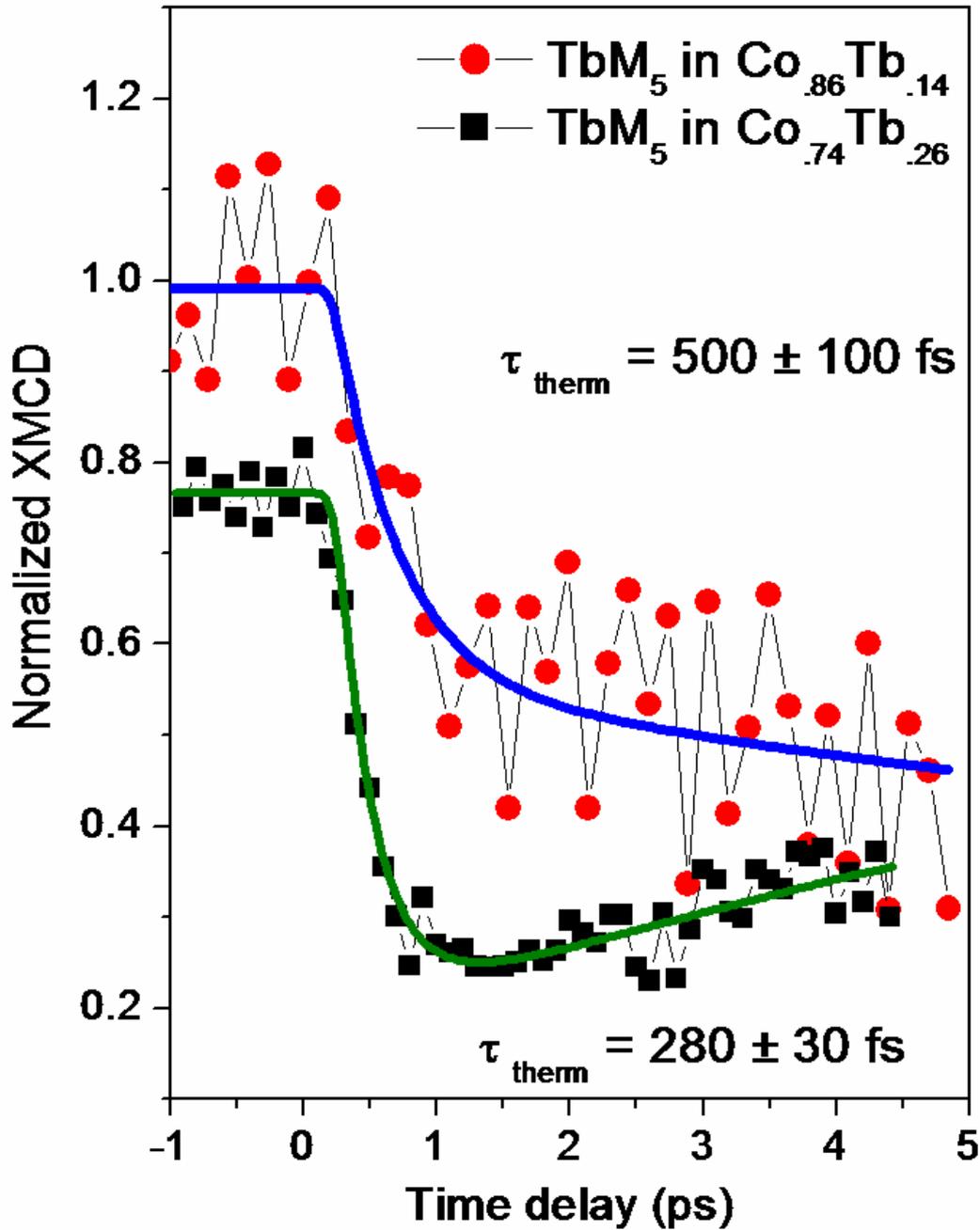

Fig 4: Lopez-Flores et al.

Figure 4 : (Color online) Ultrafast dynamics of Tb $M_5$ in $Co_{.86}Tb_{.14}$ (red circles) and $Co_{.74}Tb_{.26}$ (black squares) and the respective fit functions. The extracted thermalization time $\tau_{therm}$ obtained at the Tb $M_5$ edge in $Co_{.86}Tb_{.14}$ is $500 \pm 100$ fs, whereas in $Co_{.74}Tb_{.26}$ $\tau_{therm} = 280 \pm 30$ fs. The working temperature of the sample $Co_{.74}Tb_{.26}$ is T= 400K whereas $T_{comp}$ = 550K. The working temperature of $Co_{.86}Tb_{.14}$ is T= 400K whereas $T_C$ = 700K. During the pump-probe experiments different values of laser fluences were used (8 and resp. 21 mJ/cm$^2$), corresponding to the difference in the temperature intervals between working temperatures and the laser excited state temperatures where ultrafast demagnetization is observed.